\documentclass[%
 aip,
cp,  
 amsmath,amssymb,
 reprint,%
]{revtex4-2}

\usepackage{graphicx}
\usepackage{dcolumn}
\usepackage{bm}
\usepackage[%
    setpagesize=false,
    bookmarks=true,%
    bookmarksdepth=tocdepth,%
    bookmarksnumbered=true,%
    colorlinks=true,%
    linkcolor=,
    pdftitle={},%
    pdfsubject={},%
    pdfauthor={},%
    pdfkeywords={}%
   ]{hyperref}
\usepackage{environ}
\NewEnviron{eqsp}{%
\begin{equation}\begin{split}
  \BODY
\end{split}\end{equation}
}

\begin{document}

\preprint{AIP/123-QED}

\title{Contribution of the Weinberg-type operator to atomic electric dipole moments}

\author{Naohiro Osamura}
\email[Corresponding author: ]{osamura.naohiro.j2@s.mail.nagoya-u.ac.jp}
\affiliation{Department of Physics, Nagoya University, Furo-cho Chikusa-ku, Nagoya 464-8602 Japan}

\date{\today}

\begin{abstract}
    We report the calculation of the isovector CP-odd $\pi N$ interaction generated by the Weinberg operator, the CP violating three-gluon operator. This enables to restrict models which induce the operator by using experimental data of atomic electric dipole moment. The QCD sum rules are used to determine the matrix element required to assess the CP-odd $\pi N$ interactions. The relation between the Weinberg operator and the Hg and Xe EDMs is clarified, and the constraint on the Weinberg operator is obtained using the experimental data for the Hg EDM.
\end{abstract}

\maketitle

\section{Introduction}
    While the Standard Model (SM) has succeeded to explain much experimental data of particle physics with high precision, some experimental results suggest the existence of new physics and they are motivating us to perform active experimental searches for new physics beyond the SM (NPBSM). At exploring the NPBSM, the CP violation plays an important role and has been studied until today with the following three motivations. The first point is to realize the matter and antimatter asymmetric universe. According to Sakharov, it is known that a matter-dominated universe requires large CP violation \cite{Sakharov:1967dj}. However, the CKM phase in the SM \cite{Kobayashi:1973fv} is found to be insufficient. Secondly, from the point of the naturalness, it is possible to have CP violation in all scale. The discovery of the neutrino oscillation revealed the existence of the right-handed neutrinos \cite{Super-Kamiokande:1998kpq}. It suggests that the lepton sector also holds other CP breaking sources in the PMNS matrix at the neutrino mass scale. It naturally makes us expect that other CP breaking sources would exist in various energy scales, and new physics beyond TeV scale may violate the CP symmetry. The last point is that the electric dipole moment (EDM) which is a CP violating observable is sensitive to the NPBSM with the SM prediction being much less than the sensitivity of the EDM experiments. Therefore, once an EDM of some system is observed, this would be a direct evidence of the NPBSM \cite{Yamanaka:2015ncb}.

    Diamagnetic atom EDMs are sensitive to hadronic CP breaking sources \cite{Yamanaka:2017mef,Chupp:2017rkp}. One of the important CP breaking sources is the Weinberg operator, which is given by the following purely gluonic dimension-six operator
    \begin{equation}
        \mathcal{L}_W \equiv \frac{1}{3 !} w f^{a b c} \epsilon^{\nu \rho \alpha \beta} G_{\mu \nu}^a G_{\alpha \beta}^b G_\rho^{c \mu} .
    \end{equation}
    This operator was first calculated in the Higgs doublet model \cite{Weinberg:1989dx}, and then, it has been found that some other models also generate it such as the supersymmetric model \cite{Dai:1990xh}. While the relation between this operator and the neutron EDM has been investigated widely \cite{Demir:2002gg,Haisch:2019bml,Yamanaka:2020kjo}, its value for the CP-odd nucleon-nucleon ($N N$) interaction, which is expected to give the main contribution to the atomic EDMs, has only been evaluated quite recently \cite{Osamura:2022rak,Yamanaka:2022qlu}.
    
    In this proceedings contribution, we report on the calculation of the CP-odd $N N$ interaction generated by the isovector CP-odd $\pi N$ interaction by considering the Weinberg operator as the CP violation in the quark level. In the second section, the chiral perturbation theory will be discussed in order to show how the Weinberg operator's isovector CP-odd $\pi N$ interaction is induced. The low energy constant required to analyze the above process is calculated using QCD sum rules in the following section. Finally, we report the results of the atomic EDMs as well as the constraint on the Weinberg operator.
    
\section{Chiral perturbation theory}
    The Weinberg operator does not violate the chiral symmetry, so it is expected that the neutron EDM and the contact $N N$ interaction are the leading contribution to atomic EDMs, and the CP-odd $\pi N$ interactions are next-to-leading order (NLO) in the context of the chiral perturbation theory. However, the contribution of the contact $N N$ interaction in light nuclei is not large \cite{Yamanaka:2022qlu,Yamanaka:2015qfa} while the pion exchange contribution is enhanced in heavy nuclei, so we consider only the isovector CP-odd $\pi N$ interaction
    \begin{equation}
        \mathcal{L}_{\pi N N} \ni \sum_{N = p, n} \bar{g}^{(1)}_{\pi N N} \pi^0 \bar{N} N.
    \end{equation}
    One of the object of this proceedings contribution is to compare the neutron EDM contribution to that isovector CP-odd $\pi N$ interaction.
    
\begin{figure}
        \centering
        \includegraphics[width=0.7\linewidth]{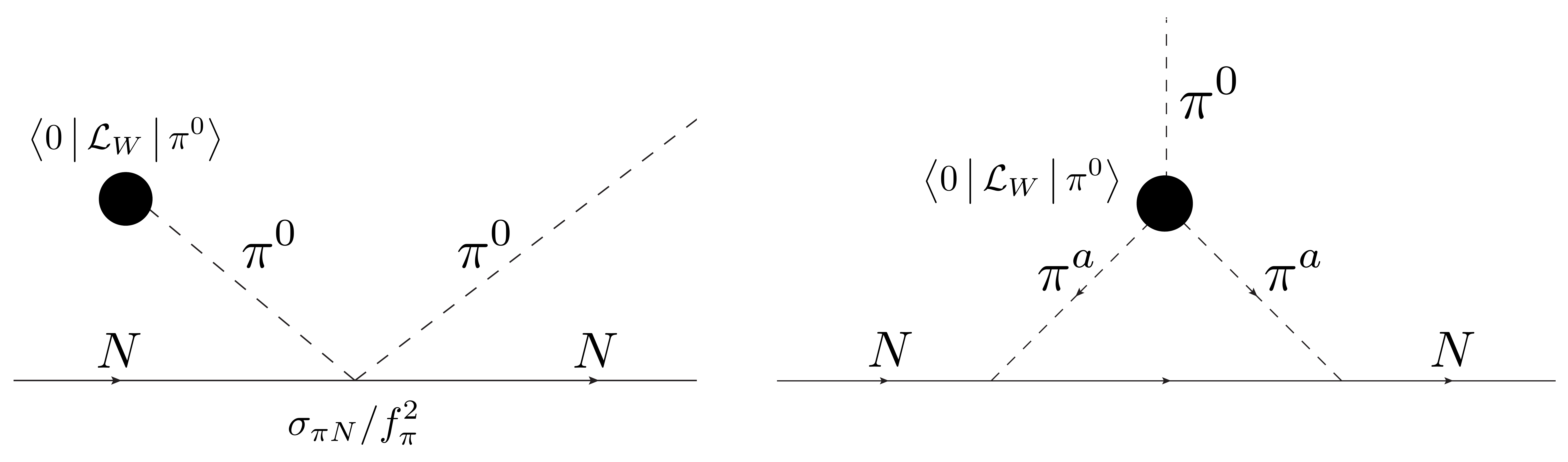}
        \caption{Diagrams generating the isovector CP-odd $\pi N$ interaction through the Weinberg operator. The left diagram corresponds to a leading contribution and the right one is taken a loop suppression and also the higher order in the chiral perturbation theory so that it is recognized as NLO.}
        \label{fig:piN interaction}
    \end{figure}

    The diagrams drawn in fig.~\ref{fig:piN interaction} contribute to $\bar{g}^{(1)}_{\pi NN}$ generated by the Weinberg operator. We need to estimate the pion-vacuum transition matrix element $\langle 0 | \mathcal{L}_W | \pi^0 \rangle$, and it will be derived in the next section using QCD sum rules. We analyse up to the one-loop level (fig.~\ref{fig:piN interaction}). The tree level contribution of $\bar{g}^{(1)}_{\pi NN}$ (see the left diagram in fig.~\ref{fig:piN interaction}) has the form
    \begin{equation}
        \bar{g}^{(1)}_{\pi N N \text{, LO}} = \langle 0 | \mathcal{L}_W | \pi^0 \rangle \frac{1}{m_{\pi}^2} \langle \pi^0 N | \mathcal{L}_{\rm QCD} | \pi^0 N \rangle .
    \end{equation}
    We already know that this chiral symmetry breaking interaction must be proportional to the quark mass, and it can be verified by noting that $\langle 0 | \mathcal{L}_W | \pi^0 \rangle \propto m_q$, $m_{\pi}^2 \propto m_q$ and $\langle \pi^0 N | \mathcal{L}_{\rm QCD} | \pi^0 N \rangle \propto m_q$. However, the four-point $\pi^0 N$ interaction, proportional to the $\pi N$ sigma term $\sigma_{\pi N}$, has also a quark-mass suppression. $\sigma_{\pi N}$ has a discrepancy between the estimations of lattice ($30$ MeV \cite{Yamanaka:2018uud,Yang:2015uis,Alexandrou:2019brg}) and phenomenology ($60$ MeV \cite{Hoferichter:2015dsa,Friedman:2019zhc}). Thus, the large value of $\sigma_{\pi N}$ enhances $\bar{g}_{\pi N N}^{(1)}$ and makes its contribution comparable to the neutron EDM.

    The atomic EDMs induced by the neutron EDM and the isovector CP-odd $\pi N$ interaction through the Schiff moment \cite{Sahoo:2018ile,Yanase:2020agg,Yanase:2020oos} are expressed as
    \begin{align}
        &d^{\rm Hg} = -6.4 \times 10^{-4} d_n-2.3 \times 10^{-4} \bar{g}_{\pi NN}^{(1)} e \, \mathrm{fm}, \\
        &d^{\rm Xe} = 1.3 \times 10^{-5} d_n-1.7 \times 10^{-5} \bar{g}_{\pi NN}^{(1)} e \, \mathrm{fm}.
    \end{align}

\section{QCD sum rules}
    We estimate the low energy constant $\langle 0 | \mathcal{L}_W | \pi^0 \rangle$ demanded in the previous section by using QCD sum rules \cite{Gubler:2018ctz}. In order to estimate the matrix element, we take the strategy to extract the matrix element as a one-particle pole in the correlation function $\langle 0 | \mathcal{L}_W (x) \mathcal{L}_W (0) | 0 \rangle$. This correlation function is expanded in terms of the intermediate states using the dispersion relation
    \begin{eqsp}
        \Pi\left(q^2\right) &=i \int d^4 x e^{-i q \cdot x}\left\langle 0\left|T\left[\mathcal{L}_W(x) \mathcal{L}_W(0)\right]\right| 0\right\rangle
        = \frac{\lambda_\pi^2}{m_\pi^2 - q^2} + \frac{1}{\pi} \int_{s_{\mathrm{th}}}^{\infty} d s \frac{\operatorname{Im} \Pi_{\mathrm{OPE}}(s)}{s-q^2},
    \end{eqsp}
    where $\lambda_\pi = \langle 0 | \mathcal{L}_W | \pi^0 \rangle$ is the desired low energy constant and $\Pi_{\rm OPE}$ is a correlation function derived for the Operator Product Expansion (OPE) calculation. To enhance the accuracy of $\lambda_\pi$, we apply the Borel transformation and differentiation with respect to $(m_u - m_d)^2 \equiv m_-^2$. The $m_-^2$ differentiation removes the glueball resonance and facilitates the extraction of the desired pion pole.  As a result, we get this correlation function for phenomenology as
    \begin{equation}
        \mathcal{B}\left[\Pi_{\text {phen }}\right]\left(M^2\right)=\lambda_\pi^2 \cdot \frac{1}{M^2} e^{-m_\pi^2 / M^2}+\frac{1}{\pi} \int_{s_{t h}}^{\infty} d s \frac{e^{-s / M^2}}{M^2} \operatorname{Im} \Pi_{\mathrm{OPE}}\left(s ; m_{-} \neq 0\right)
    \end{equation}
    This step leaves two undetermined parameters; the Borel mass $M$ and the effective threshold parameter $s_{\rm th}$.

\begin{figure}
        \begin{minipage}[b]{0.24\linewidth}
            \centering
            \includegraphics[width=1\linewidth]{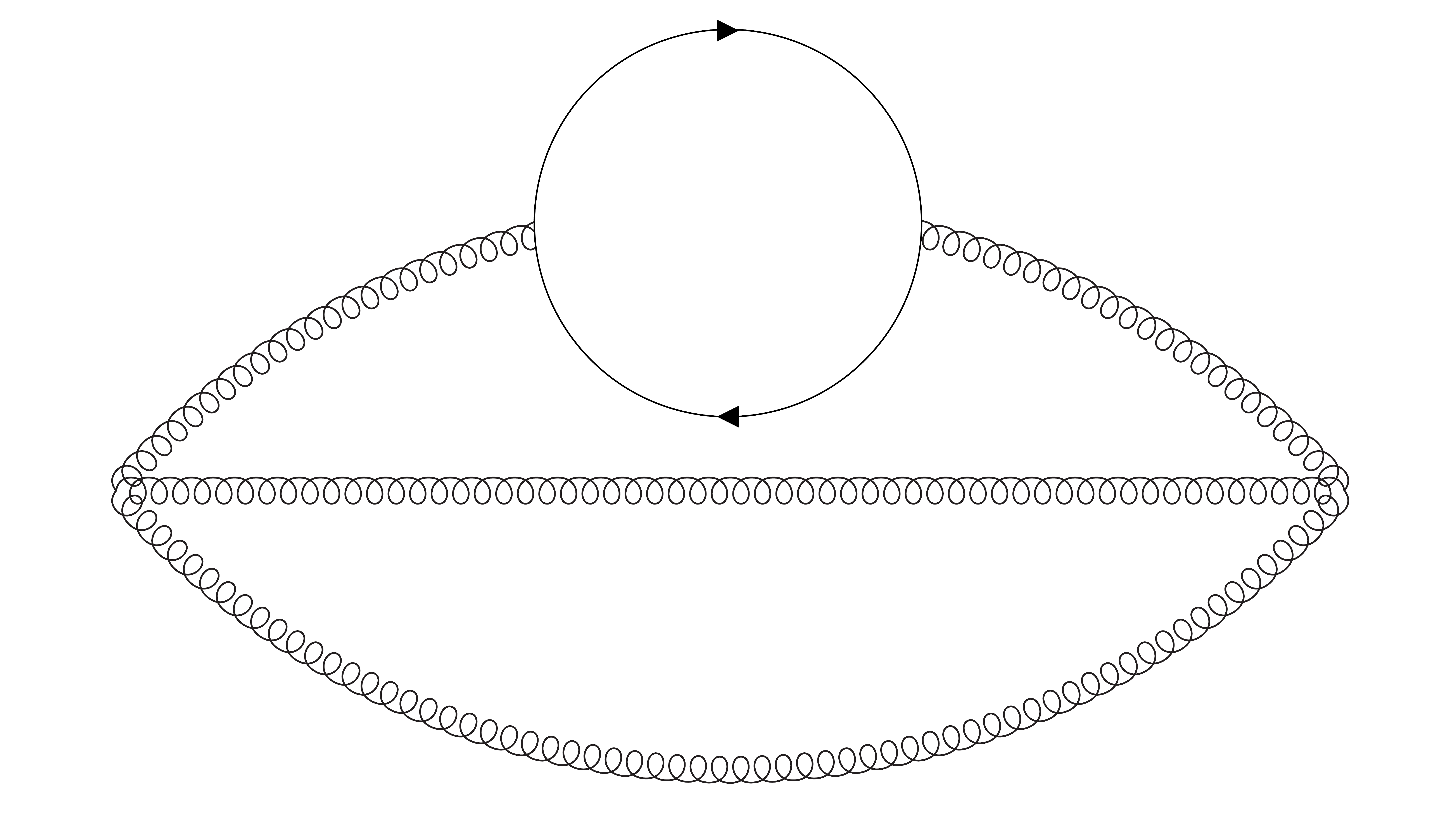}
            \label{fig-loop}
        \end{minipage}
        \begin{minipage}[b]{0.24\linewidth}
            \centering
            \includegraphics[width=1\linewidth]{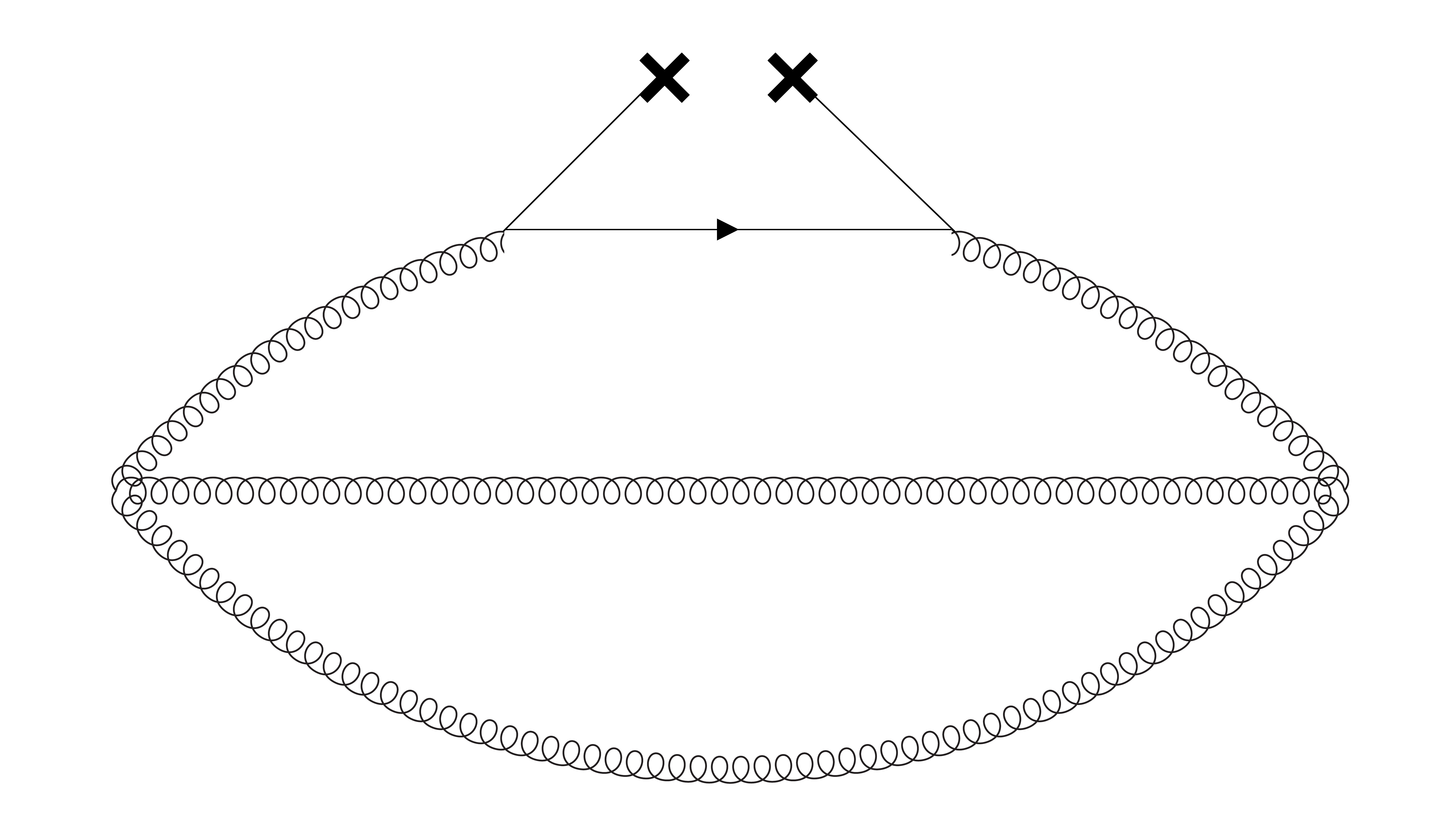}
            \label{fig-q}
        \end{minipage}
        \begin{minipage}[b]{0.24\linewidth}
            \centering
            \includegraphics[width=1\linewidth]{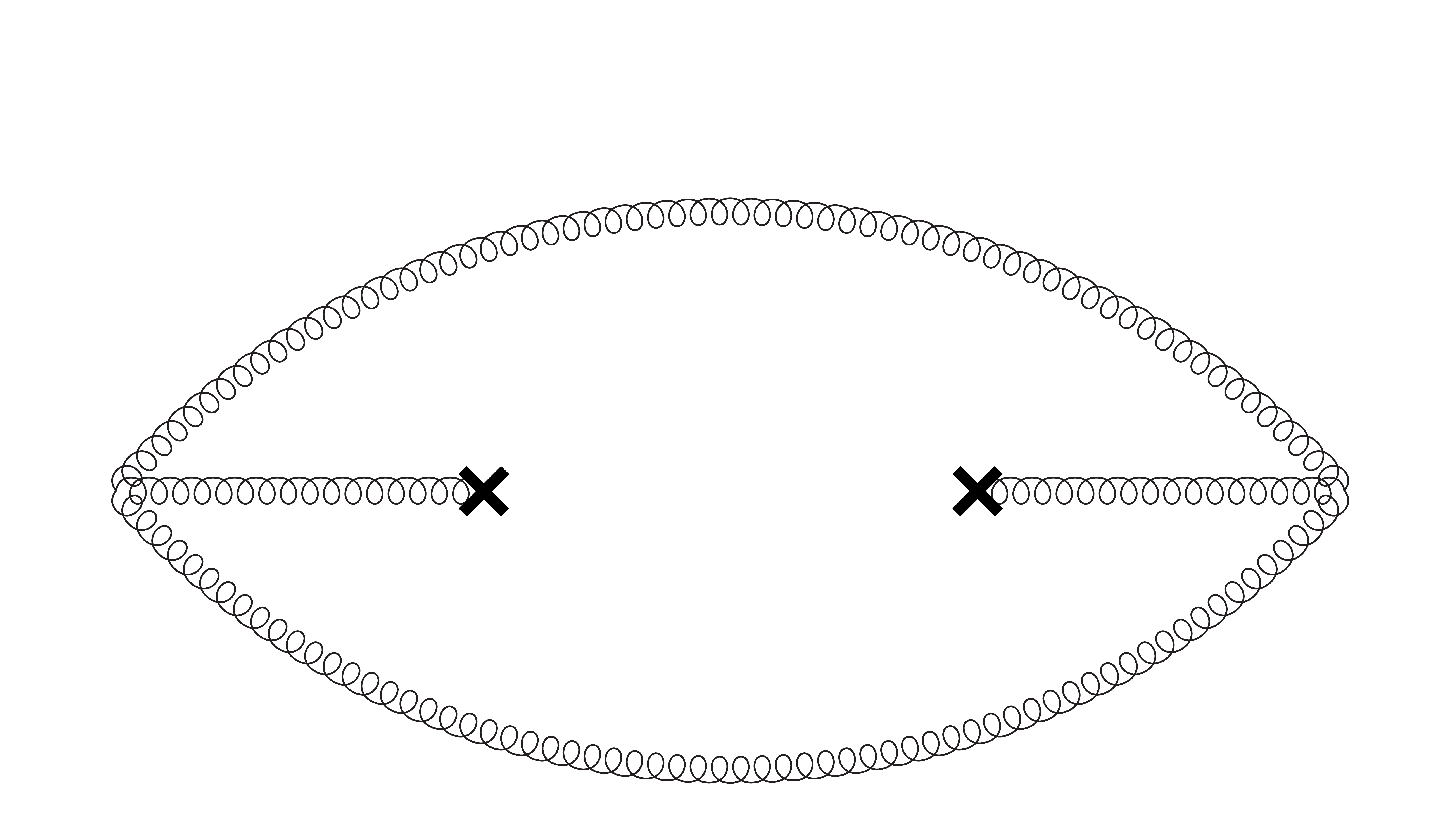}
            \label{fig-G}
        \end{minipage}
        \begin{minipage}[b]{0.24\linewidth}
            \centering
            \includegraphics[width=1\linewidth]{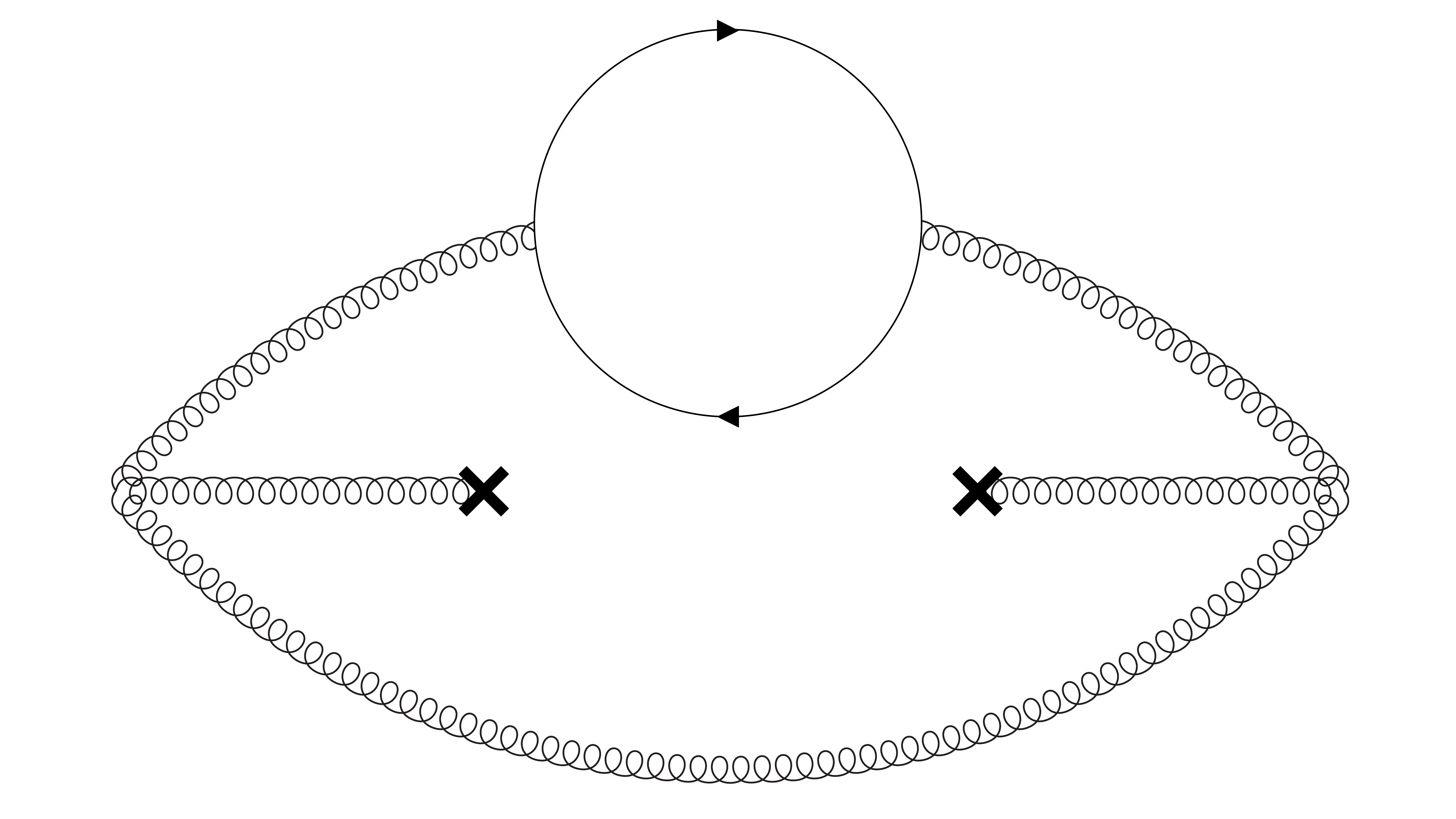}
            \label{fig-loop+G}
        \end{minipage}
        \caption{The diagrams contributing to $\Pi_{\text{OPE}} (m_-^2 \neq 0)$. The crosses in the latter three diagrams represent the chiral condensate $\langle 0|\bar{q}q|0 \rangle$, and the gluon condensate $\langle 0|\alpha_s G^2|0 \rangle$. The two diagrams on the right, which both include a gluon condensate contribution, vanish in the OPE calculation of this work.}
        \label{fig-diagrams}
    \end{figure}

    Now let us do the OPE calculation. The OPE enables us to compute the correlation function having the non-perturbative effect of QCD. We split $\Pi_{\rm OPE}$ into the isospin symmetric part and the isospin breaking one. The isospin symmetric part $\Pi_{\rm OPE} (m_-^2 = 0)$ vanishes by the $m_-^2$ differentiation. Thus, we just focus on the isospin breaking part $\Pi_{\rm OPE} (m_-^2 \neq 0)$, which is evaluated by a few lowest dimension operators (see fig.~\ref{fig-diagrams}). The correlation function for OPE, after the Borel transformation, is obtained as
    \begin{equation}
        \mathcal{B}\left[\Pi_{\mathrm{OPE}}\left(q^2 ; m_{-} \neq 0\right)\right]=w^2\left[-\frac{3 m_{-}^2 \alpha_s}{64 \pi^4} M^6+\frac{m_{-}^2 B_0^2 h_3 \alpha_s}{2 \pi^3} M^4\right] .
    \end{equation}
    As a result of the whole sum rules calculation, we obtain
    \begin{equation}
        \left|\left\langle 0\left|\mathcal{L}_W\right| \pi^0\right\rangle\right| \in w \cdot[2.3,8.3] \times 10^{-6} \mathrm{GeV}^5 .
    \end{equation}
    Here, the theoretical uncertainty is estimated from the variation of the Borel mass $M$ and the threshold parameter $s_{\rm th}$.

\section{Numerical analysis}
    We finally show the results of the atomic EDMs generated by the Weinberg operator:
    \begin{align}
        &d^{\rm Hg} =w \left(- 1.3 \times 10^{-2} \pm [0.71,4.4] \times 10^{-3}  \right) e \text{ MeV}, \\
        &d^{\rm Xe} = w \left( 2.7 \times 10^{-4} \pm [0.52,3.2] \times 10^{-4} \right) e \, {\rm MeV} ,
    \end{align}
    where the first terms correspond to the neutron EDM contribution and the second terms are from the isovector CP-odd $\pi N$ interaction calculated in this work \cite{Osamura:2022rak}. Here, the two numbers in the squared brackets represent the theoretical uncertainty.
    This revealed that the $\pi N$ interaction should not be neglected; in Hg EDM, the ratio between the maximal contribution of the $\pi N$ interaction and that of the neutron EDM is 34\%. The remarkable result is that of Xe. The $\pi N$ interaction may exceed that of the neutron EDM. Then, we derive the constraint on the Weinberg operator by using the experimental data of the Hg EDM \cite{Graner:2016ses}
    \begin{equation}
        |w(\mu=1 \mathrm{TeV})|<4 \times 10^{-10} \mathrm{GeV}^{-2}.
    \end{equation}

\section{Conclusion}
    We reported the calculation of the relation between the Weinberg operator and the atomic EDM through the CP-odd $\pi N$ interaction. It has been anticipated that the contribution of the latter is much less than that of the neutron EDM according to chiral perturbation. However, this study showed that the $\pi N$ interaction should not be neglected. In particular, for the Xe EDM, the contribution of the $\pi N$ interaction may exceed that of the neutron EDM.

\end{document}